\documentstyle[preprint,aps]{revtex}

\begin{document}
\draft
\title{Motion of a heavy impurity through a Bose-Einstein condensate.}
\author{G.E.~Astrakharchik$^{ab}$, L.P.~Pitaevskii$^{ac}$}
\address{$^{a}$Dipartimento di Fisica, Universit\`{a} di Trento and Istituto\\
Nazionale per la Fisica della Materia, I-38050 Povo, Trento, Italy;\\
$^{b}$Institute of Spectroscopy,142190 Troitsk, Moscow region, Russia;\\
$^{c}$Kapitza Institute for Physical Problems, 119334 Moscow, Russia}
\maketitle

\begin{abstract}
\noindent We study motion of a point-like impurity in a Bose-Einstein
condensate at $T=0$. By solving the Gross-Pitaevskii equation in a perturbative manner we
calculate the induced mass of the impurity and the drag force on the
impurity in 3D, 2D and 1D cases. The relationship between the induced mass
and the normal mass of fluid is found and coincides with the result of the
Bogoliubov theory. The drag force appears for supersonic motion of the
impurity. In 1D the drag force is investigated also on the basis of the
exact Lieb-Liniger theory, using the dynamic form factor, which has been
evaluated by the Haldane method of the calculation of correlation functions.
In this theory the force appears for an arbitrarily small velocity of the
impurity. The possibility of measuring the form factor in existing
experiments is noted.
\end{abstract}


\section{Introduction}

One of the most important peculiarities of Landau theory superfluidity is
the existence of a finite critical velocity. If \ a body moves in a
superfluid at $T=0$\ with velocity $V$\ \ less then $v_{c}$, the motion is
dissipationless. At $V>v_{c}$\ a drag force arises because of the
possibility of emission of elementary excitations. However, both theoretical
and experimental investigation in superfluid $^{4}He$\ are difficult. The
critical velocity in $^{4}He$\ is related to creation of rotons, for which
one has no simple theoretical description. Further, an important role is
played by complicated processes involving vortex rings production.

The situation in low-density weakly-interacting Bose-Einstein condensed
(BEC) gases is simpler. The Landau critical velocity in this case is due to
Cherenkov emission of phonons which can be described by mean-field theory.
Due to the presence in the theory of an intrinsic length parameter - the
correlation length $\xi $- the friction force for a small body does not
depend on its structure. Vortex rings in the BEC \ cannot have radius less
then $\xi $\ \ (see \cite{Rob}) and it is reasonable to believe that
probability of their creation by a small body is small. Thus quantitative
investigation of critical velocities in BEC are very interesting and can be
used to probe the superfluidity of a quantum gas.

Recently existence of the critical velocity in a Bose-Einstein Condensed gas
was confirmed in a few experiments. At MIT a trapped condensate was stirred
by a blue detuned laser beam \cite{Raman et al} and the energy of
dissipation was measured. The critical velocity was found to be smaller than
the speed of sound due to emission of vortices. The diameter of the laser
spot in this experiment was of a macroscopic size and was large compared to
the healing length. An improved technique allowed measurement of the drag
force acting on the condensate in a subsequent experiment \cite{Onofrio et
al}.

The analytical study of flow of the condensate over an impurity is highly
nontrivial due to the intrinsic nonlinearity of the problem arising from the
interaction of the particles in the condensate. In one dimension the
dissipation could occur at velocities smaller than predicted by Landau's
approach due to emission of solitons \cite{Hakim}. The dependence of the
critical velocity on the type of the potential was studied both by using a
perturbative approach and numerical integration in \cite{Leboeuf et
al,Pavloff}. The effective two dimensional problem was considered in \cite
{Kagan}. In this work generation of excitations in the oscillating
condensate in a time dependent parabolic trap in the presence of a static
impurity was studied analytically. A three-dimensional flow of a condensate
around an obstacle was calculated numerically by integration of the
Gross-Pitaevskii (GP) equation and emission of vortices was observed \cite
{Frisch et al,Winiecki et al}.

In this letter we study a $\delta $-function perturbation moving at a
constant velocity in a condensate. We find analytically the depletion of the
superfluid fraction and the drag force.

\section{Three-dimensional system \label{3D}}

Let us consider an impurity moving through a three-dimensional condensate at
$T=0$. One of the possible realizations of this model could be scattering of
heavy neutral molecules by the condensate.

\subsection{Effective mass and normal fraction \label{total energy}}

We start from the three-dimensional energy functional of a homogeneous
weakly-interacting Bose gas in the presence of a $\delta $-function
perturbation (an impurity) moving with a constant velocity ${\bf V}$

\begin{equation}
E\!=\!\!\int \!\!\left( \frac{\hbar ^{2}}{2m}|\nabla \psi |^{2}\!+\!(\mu
\!-\!g_{i}\delta ({\!}{\bf r}-\!{\bf V}t))|\psi |^{2}\!+\!\frac{g}{2}|\psi
|^{4}\!\right) \!d^{3}x,  \label{energy functional}
\end{equation}
where $\psi $ is the condensate wavefunction, $\mu $ is the chemical
potential, $m$ mass of a particle in the condensate, $g=4\pi \hbar ^{2}a/m$
and $g_{i}=2\pi \hbar ^{2}b/m$ are particle-particle and particle-impurity
coupling constants, with $a$ and $b$ being the respective scattering lengths
\cite{hom}. We will assume that the interaction with impurity is small and
we will use perturbation theory. By splitting the wave function into a sum
of the unperturbed solution and a small correction $\psi ({\bf r},t)=\phi
_{0}+\delta \psi ({\bf r},t)$ and linearizing the time-dependent GP equation
with respect to $\delta \psi $, we obtain an equation describing the time
evolution of $\delta \psi $

\begin{eqnarray}
i\hbar \frac{\partial }{\partial t}\delta \psi &=&\left( -\frac{\hbar ^{2}}{%
2m}\triangle -\mu +2g|\phi _{0}|^{2}\right) \delta \psi +g\phi
_{0}^{2}\,\delta \psi ^{\ast }  \nonumber \\
&&+g_{i}\,\delta ({\bf r}-{\bf V}t)\phi _{0}\ .  \label{GPE delta Psi}
\end{eqnarray}

\noindent In a homogeneous system $\phi _{0}$ is a constant fixed by the
particle density $\phi _{0}=\sqrt{n}$ and $\mu =gn=mc^{2}$.

The perturbation follows the moving impurity, i.e. $\delta \psi $ is a
function of $({\bf r}-{\bf V}t)$, so the coordinate derivative is related to
the time derivative $\partial \delta \psi ({\bf r}-{\bf V}t)/\partial t=-%
{\bf V}\vec{\nabla}\delta \psi ({\bf r}-{\bf V}t)$. We shall work in the
frame moving with the impurity ${\bf r}^{\prime }={\bf r}-{\bf V}t$ and the
subscript over ${\bf r}$ will be dropped.

Eq. (\ref{GPE delta Psi}) for a perturbation in a homogeneous system can be
conveniently solved in momentum space. In order to do this we introduce the
Fourier transform of the wave function $\delta \psi _{{\bf k}}=\int e^{-i%
{\bf k\cdot }{\bf r}}\delta \psi ({\bf r})\,d^{3}x$. Eq. (\ref{GPE delta Psi}%
) becomes

\begin{equation}
\left(\hbar {\bf k\!\cdot\! V}\!-\!\frac{\hbar ^{2}k^{2}}{2m}%
\!-\!mc^{2}\right)\delta \psi _{{\bf k}}\!-\!mc^{2}(\delta \psi _{{\bf -k}%
})^{\ast }\!-\!g_{i}\phi _{0}\!=\!0.  \label{GPE delta Psi fourier}
\end{equation}

Substitution of $k\rightarrow -k$ and complex conjugation of (\ref{GPE delta
Psi fourier}) give the second equation. The obtained system of linear
equations can be easily solved

\begin{equation}
\delta \psi _{{\bf k}}=g_{i}\phi _{0}\frac{\hbar {\bf k\cdot V}+\frac{\hbar
^{2}k^{2}}{2m}}{(\hbar {\bf k\cdot V})^{2}-\frac{\hbar ^{2}k^{2}}{2m}\left(
\frac{\hbar ^{2}k^{2}}{2m}+2mc^{2}\right) }\ .  \label{dPsi}
\end{equation}

Let us calculate the energy of the perturbation. By neglecting terms of the
order of $\ g_{i}|\delta \psi |^{2}$ the energy functional (\ref{energy
functional}) becomes

\begin{equation}
E\!=\!E_{0}\!+\!g_{i}\phi _{0}^{2}\!+\!\int \!\!\hbar {\bf k\cdot V}|\delta
\psi _{{\bf k}}|^{2}\!\frac{d^{3}k}{(2\pi )^{3}}\!+\!\frac{g_{i}\phi _{0}}{2}%
(\delta \psi \!+\!\delta \psi ^{\ast })_{{\bf r}=0}.  \label{E Psi}
\end{equation}

Here $E_{0}=Ngn/2$ is the energy of the system in absence of the
perturbation. We expand the $3^{rd}$ and $4^{th}$ terms in powers of $V$. \
To avoid the large-$k$ divergency, one must also introduce a renormalization
of the scattering amplitude. It is sufficient to express the coupling
constant in the $2^{nd}$ term of eq. (\ref{E Psi}) in terms of the
scattering amplitude $b$ using the second order Born approximation:

\begin{equation}
g_{i}=\frac{2\pi \hbar ^{2}b}{m}\left( 1+\frac{2\pi \hbar ^{2}b}{m}\int
\left( \frac{\hbar ^{2}k^{2}}{2m}\right) ^{-1}\frac{d^{3}k}{(2\pi )^{3}}%
\right) \ .
\end{equation}

Finally, by carrying out the integration over momentum space and considering
$N_{imp}$ impurities with a concentration given by $\chi =N_{imp}/N$ we
obtain the energy per particle

\begin{eqnarray}
\frac{E}{N} &=&\left\{2\pi na^{3}\left( 1+\chi \frac{b}{a}\right) +8\pi
^{3/2}(na^{3})^{3/2}\chi \left( \frac{b}{a}\right) ^{2}\right\} \frac{\hbar
^{2}}{ma^{2}}  \nonumber \\
&&+\frac{2\sqrt{\pi }}{3}(na^{3})^{1/2}\chi \left( \frac{b}{a}\right) ^{2}%
\frac{mV^{2}}{2}\ .  \label{E}
\end{eqnarray}

If we set $V=0$ we recover Bogoliubov's corrections to the energy in the
presence of quenched impurities \cite{Huang,Disorder}. Note that even if the
``mean-field'' energy obtained from the GP equation in the absence of
impurities ($\chi =0$) \ leaves out terms of the order of $(na^{3})^{3/2}$,
the equations we obtain in the presence of impurities in a perturbative
manner still correctly describe the effect of the disorder up to the terms
of the order of $(na^{3})^{3/2}$.

If $V\neq 0$ a quadratic term in the impurity contribution to the energy is
present. It can be denoted as $\chi m^{\ast }V^{2}/2$ with
\begin{equation}
m^{\ast }=\frac{2\sqrt{\pi }}{3}(na^{3})^{1/2}\left( \frac{b}{a}\right) ^{2}m
\label{mstar}
\end{equation}
being the induced mass, i. e. the mass of particles dragged by an impurity
\cite{Thesis}. \ Applicability of the perturbation theory demands $m^{\ast }$
to be small compared to $m.$ This gives the condition $(na^{3})^{1/2}\left(
\frac{b}{a}\right) ^{2}\ll 1$.\ At zero temperature the interaction between
particles does not lead to depletion of the superfluid density and the
suppression of the superfluidity comes only from the interaction of
particles with impurities. Thus $\left( \ref{mstar}\right) $ defines the
normal density $\rho _{n}=\chi \rho m^{\ast }/m$.

This result is in agreement with the one obtained by the means of Bogoliubov
transformation starting from the Hamiltonian written in the second-quantized
form in the presence of disorder \cite{Huang,Disorder}.

The normal density of a superfluid is an observable quantity. It was
evaluated in liquid $^{4}He$\ by measuring of the moment of inertia of a
rotating liquid or by measuring of the second sound velocity. Both methods
can be, in principle, developed for BEC gases.

\subsection{Drag force and energy dissipation \label{energy dissipation}}

The force with which the impurity acts on the system is

\begin{equation}
{\bf F}=-\int |\psi \left( {\bf r}\right) |^{2}\,\vec{\nabla}(g_{i}\delta
\left( {\bf r}\right) )\,d^{3}x=g_{i}(\vec{\nabla}|\psi({\bf r})|^{2})_{{\bf %
r}=0}\ .
\end{equation}

Expanding the wave function into the sum of $\phi _{0}$ and $\delta \psi $
and neglecting terms of order $\delta \psi ^{2}$ we obtain

\begin{eqnarray}
{\bf F} &=&g_{i}\phi _{0}\int i{\bf k}\left[ \delta \psi _{{\bf k}}+\left(
\delta \psi _{{\bf -k}}\right) ^{\ast }\right] \frac{d^{3}k}{(2\pi )^{3}}
\nonumber \\
&=&\int \frac{2\left( g_{i}\phi _{0}\right) ^{2}i{\bf k}\left( \hbar
^{2}k^{2}/2m\right) }{(\hbar {\bf k\cdot V}+i0)^{2}-\frac{\hbar ^{2}k^{2}}{2m%
}\left( \frac{\hbar ^{2}k^{2}}{2m}+2mc^{2}\right) }\frac{d^{3}k}{\left( 2\pi
\right) ^{3}}\ ,  \label{force}
\end{eqnarray}
where we added an infinitesimal positive imaginary part $+i0$\ to the
frequency ${\bf k\cdot V}$\ according to the usual Landau causality rule.
The drag force is obviously directed along to the velocity ${\bf V}$.{\bf \ }

We can carry out the integration with respect to $\cos \vartheta ,$ where $%
\vartheta $ is the angle between the momentum ${\bf k}$ and velocity ${\bf V}
$, using the formula $\frac{1}{x+i0}={\cal P}\frac{1}{x}-i\pi \delta (x)$.
Due to the integration between symmetric limits, only the imaginary part
contributes to the integral with respect to $\cos \vartheta $. The poles in
the integration over $\cos \vartheta $ appear if the square root in the
denominator is smaller than one, which leads to the restriction on the
values of momentum which contribute

\begin{equation}
|k| \leq k_{\max} = 2m (V^2-c^2)^{1/2}/\hbar\ .  \label{kmax}
\end{equation}
Thus the energy dissipation takes place only if the impurity moves with a
speed larger than the speed of sound. Integration with respect to $k$,
taking into account restriction (\ref{kmax}), finally gives

\begin{equation}
F_{V}=4\pi nb^{2}mV^{2}(1-c^{2}/V^{2})^{2}\ .  \label{F3D}
\end{equation}
The energy dissipation, \ $\dot{E}=-F_{V}V$, can be evaluated by measuring
the heating of the gas.\

For large $V$ the force is proportional to $\ V^{2}$. The energy dissipation
per unit time can then be presented as $\dot{E}=-\gamma E$ with the damping
rate $\gamma \sim nb^{2}V.$

Note in conclusion that our perturbative calculations can not describe
processes involving dissipation of energy due to creation of quantized
vortex rings. Such a creation is possible at $V<c$ but has a small
probability for low velocity and for a weak point-like impurity.

\section{Low dimensional systems \label{low dimensional systems}}

In this type of experiment the role of the impurity can also be played by a
laser beam with small enough size and intensity. The Fourier components of
the perturbed wavefunction $\delta \psi_{{\bf k}}$ are given by the formula (%
\ref{dPsi}), which is derived in an arbitrary number of dimensions. The only
difference is in the substitution of $d^3k/(2\pi)^3$ with $d^D\!k/(2\pi)^D$
in the integrals.

\subsection{Two-dimensional system \label{2D}}

There are different possible geometries of the experiment. One can create a
two-dimensional perturbation in the three-dimensional condensate. Such a
two-dimensional impurity can be created, analogously to the MIT experiment
\cite{Raman et al,Onofrio et al}, by a thin laser beam. Such a beam creates
a cylindrical hole in the condensate, which is stirred by moving the
position of the laser beam. Another possibility is to fix the position of
the laser beam along the long axis of an elongated condensate, so that the
dissipation can be studied by shaking the trap and exciting the breathing
modes. The problem is to create a beam with a diameter which is small with
compared to the correlation length. The theory can be easily generalized for
beams of finite diameter. The intensity of the beam can be tuned to satisfy
the condition of a weak perturbation.

The more interesting possibility is the investigation of true
two-dimensional condensates, which can be created in plane optical traps,
produced by a standing light wave. If the light intensity is large enough,
tunneling between planes is small and the condensates behave as independent
two dimensional systems. The impurity can again be created by a laser beam
perpendicular to the condensate plane. Another possibility is to use
impurity atoms, which can be drive by a laser beam, with a frequency close
to the atomic resonance of the impurity.

In two dimensions one has $d^{2}k=kdkd\vartheta $. The drag force ${\bf F}$
is different from zero only if the denominator has poles, which means that
the velocity $V$ must be larger than the speed of sound $c$. Only momenta
smaller than $k_{max}$ (see eq. (\ref{kmax})) contribute to the integral.

Integration for the velocities $V>c$ gives us the drag force in the
two-dimensional case:
\begin{equation}
F_{V}^{2D}=g_{i}^{2}n_{2}m^{2}(V^{2}-c^{2})/(\hbar ^{3}V)\ ,  \label{FV2}
\end{equation}
where $n_{2}$ is two-dimensional density.

In a quasi two-dimensional system, i.e. when the gas is confined in the $z$%
-direction by the harmonic potential $\left( 1/2\right) m\omega
_{z}^{2}r^{2} $, the two-dimensional coupling constant equals $g_{i}=\sqrt{%
2\pi }\frac{\hbar ^{2}b}{ma_{z}}$, where $a_{z}=\sqrt{\hbar /m\omega _{z}}$
\ is the oscillator length and $b$\ is the three-dimensional scattering
length. (We consider here only the mean-field $2D$ situation. See \cite{PS},
\S\ 17 for a more detail discussion.)

The calculation of the effective mass gives

\begin{equation}
m^{\ast }=g_{i}^{2}n_{2}m/(4\pi ^{2}\hbar ^{2}c^{2})\ .  \label{mstar2}
\end{equation}
Notice again that our calculations do not take into account creation of
vortex pairs which is possible at $V<c$.

\subsection{One-dimensional system. Mean-field theory \label{1D}}

In one dimension the integration is straightforward. The integration $(\ref
{force})$ over $k$ gives $2\pi i$ if $V>c$ and zero otherwise. So, the force
is $F_{V}^{1D}=2g_{i}^{2}n_{1}m/\hbar ^{2}$, where $n_{1}$ is the linear
density. In a quasi one dimensional system (i.e. a very elongated trap or a
waveguide) there are no excitations in the radial harmonic confinement and
the coupling constant is given by $g_{i}=-\frac{\hbar ^{2}}{mb_{1D}}$ with $%
b_{1D}=-a_{\perp }^{2}/b$, where $a_{\perp }=\sqrt{\hbar /m\omega _{\perp }}$
and one has $F^{1D}= 2n_{1}\hbar ^{2}/mb_{1D}^{2}$.

An interesting peculiarity is that the result does not depend on the
velocity $V$ (where, of course, the velocity must be larger than the speed
of sound). The same result for the $\delta $-potential was found in \cite
{Pavloff}. Calculation of the effective mass gives $m^{\ast
}=g_{i}^{2}n_{1}/2\hbar c^{3}$.

In a 1D system energy dissipation is possible at $V<c$ due to creation of
the ``gray solitons'' first considered in \cite{Tsuz}. Non-linear
calculations \cite{Hakim} show that the critical velocity for this process
decreases with increasing coupling constant $g_{i}$.

This theory can be checked in an experiment in a three-dimensional
condensate. The two-dimensional impurity can be presented by a moving light
sheet.

\subsection{One-dimensional system. Bethe-ansatz theory}

We saw in the previous subsection that for a weakly interacting impurity the
drag force appears only when the impurity velocity $V$ is larger that the
Landau critical velocity, which is equal to the velocity of sound $c$. The
situation is, however, different in the Bethe-ansatz Lieb-Liniger theory of
a 1D Bose gas \cite{Lieb1}. According to this theory excitations in the
system actually have a fermionic nature. Even a low frequency perturbation
can create a particle-hole pair with a total momentum near $2p_{F}\equiv
2\hbar k_{F}=\hbar 2\pi n_{1}$. To calculate the drag force for this case we
will use the dynamic form factor of the system $\sigma \left( \omega
,k\right) $ (we follow notation of \cite{LP9}, \S\ 87). The dissipated
energy at $T=0$ can be calculated as
\begin{equation}
\dot{E}=-\int\limits_{-\infty }^{\infty }\frac{dk}{2\pi }\int\limits_{0}^{%
\infty }\frac{d\omega }{\pi }\omega \frac{n_{1}}{2\hbar }\sigma \left(
\omega ,k\right) \left| U\left( \omega ,k\right) \right| ^{2}\ ,  \label{Q}
\end{equation}
where $U(\omega ,k)=2\pi g_{i}\delta (\omega -kV)$ is the Fourier transform
of the impurity potential{\bf \ }$U(t,z)=g_{i}\delta (z-Vt)$. One has $%
\left| U\left( \omega ,k\right) \right| ^{2}=2\pi g_{i}^{2}t\delta (\omega
-kV)$, where $t$ is ''time of observation''. Thus the energy dissipation per
unit of time is
\begin{equation}
\dot{E}=-F_{V}V=-\frac{g_{i}^{2}n_{1}V}{\hbar }\int\limits_{0}^{\infty }%
\frac{dk}{2\pi }k\sigma \left( kV,k\right) ,  \label{F}
\end{equation}
where $F_{V}$ is the drag force. We will try to estimate the velocity
dependence of $F_{V}$.

For low frequency dissipation the important values of $k$ are near $2k_{F}$.
According to \cite{Castro}

\begin{equation}
\sigma \left( \omega ,2k_{F}\right) \sim \omega ^{\left( \eta -2\right)
},\omega \rightarrow 0,  \label{omega}
\end{equation}
where $\eta =\frac{2\hbar k_{F}}{mc}=\frac{2\pi \hbar n_{1}}{mc}\geq 2$ is
the characteristic parameter of a 1D Bose gas. In the mean-field limit when $%
n_{1}\rightarrow \infty $ the parameter $\eta \rightarrow \infty $. In the
opposite case of a small density bosons behave as impenetrable particles
(Girardeau limit \cite{Gir}) and the dynamic form-factor coincides with the
one of an ideal Fermi gas. In this limit $\eta =2.$

In the general case one can calculate $\sigma \left( \omega ,k\right) $\ at
small $\omega $\ and $\ k\approx 2k_{F}$generalizing the method of Haldane
\cite{Hald} for the case of time-dependent correlation functions.
Calculations give
\begin{equation}
\sigma \left( \omega ,k\right) =\frac{n_{1}c}{\omega ^{2}}\left( \frac{\hbar
\omega }{mc^{2}}\right) ^{\eta }f\left( \frac{c\Delta k}{\omega }\right) \
,\omega >0,k>0,  \label{So}
\end{equation}
where $k=2k_{F}+\Delta k$\ and the function $f(x)$\ is
\begin{equation}
f\left( x\right) =A\left( \eta \right) \left( 1-x^{2}\right) ^{\eta /2-1}
\end{equation}
in the interval $\ \left| x\right| <1$\ and is equal to zero at $\left|
x\right| \geq 1$\ (see also \cite{KBI}). The constant $A\left( \eta \right)$
can be calculated in two limiting cases: $A\left( \eta =2\right) =\pi /4$
(see \cite{PS} \S\ 17.3) and $A\left( \eta \right) \approx 8\pi ^{2}/\left[
\left( 8C\right) ^{\eta }\Gamma ^{2}\left( \frac{\eta }{2}\right) \right]$
, where $C=1.78$... is the Euler 's constant, for $\eta \gg 1$ (details of
the calculation will be published elsewhere).

Substituting $\left( \ref{So}\right) $\ into $\left( \ref{F}\right) $\ we
finally find velocity dependence of the drag force:
\begin{equation}
F_{V}=\frac{\Gamma \left( \frac{\eta }{2}\right) }{2\sqrt{\pi }\Gamma \left(
\frac{\eta +1}{2}\right) }A\left( \eta \right) \frac{g_{i}^{2}n_{1}^{2}}{%
\hbar V}\left( \eta \frac{V}{c}\right) ^{\eta }\;.  \label{FV}
\end{equation}
Equation $\left( \ref{FV}\right) $ is valid for the condition $V\ll c$.

Thus in the Girardeau strong-interaction limit $F_{V}\sim V$ and Bose gas
behaves, from the point of view of friction, as a normal system, where the
drag force is proportional to the velocity. On the contrary, in the
mean-field limit the force is very small and the behavior of the system is
analogous to a 3D superfluid. However, even in this limit the presence of
the small force makes a great difference. Let us imagine that our system is
twisted into a ring, and that the impurity rotates around the ring with a
small angular velocity. If the system is superfluid in the usual sense of
the word, the superfluid part must stay at rest. Presence of the drag force
means that equilibrium will be reached only when the gas as a whole rotates
with the same angular velocity. From this point of view the superfluid part
of the 1D Bose gas is equal to zero even at $T=0.$ Notice that in an earlier
paper \cite{Son} the author concluded that $\rho _{s}=\rho $ at $T=0$ for
arbitrary $\eta $. We believe that this difference results from different
definitions of $\rho _{s}$ and reflects the non-standard nature of the
system.

Equation $\left( \ref{FV}\right) $ is equivalent to a result which was
obtained by a different method in \cite{Buch}, with a model consisting of an
impurity considered as a Josephson junction. Notice that the process of
dissipation, which in the language of fermionic excitations can be described
as creation of a particle-hole pair, corresponds in the mean-field limit to
creation of a phonon and a small-energy soliton. It seems that such a
process cannot be described in the mean-field approach in the linear
approximation.

Experimental confirmation of these quite non-trivial predictions demands a
true one-dimensional condensate, where non mean-field effects can be
sufficiently large. Such condensates have been investigated for the first
time in experiments \cite{1D1,1D2}. In experiments \cite{1DA,1DB}
condensates have been created in the form of elongated independent
``needles'' in optical traps, consisting of two perpendicular standing laser
waves. The role of an impurity in this case must be played by a light sheet,
perpendicular to the axis of condensates and moving along them.

Notice also, that application of the additional light waves in this
experiments of this type allows one to create a harmonic perturbation of the
form
\begin{equation}
U(t,z)=U_{0}\cos \left( \omega t-kz\right) ,k=2k_{F}+\Delta k
\end{equation}
with small $\omega $\ and \ $\Delta k$. Such potential with was used in \cite
{1DA,1DB} \ for experiments with 1D condensate in a periodic lattice.
However, for a small amplitude $U_{0}$, measurement of the dissipation
energy $Q$\ gives, according to $\left( \ref{Q}\right) ,$\ the dynamic
form-factor $S\left( \omega ,k\right) $ directly.

\section{Conclusions \label{conclusions}}

We have studied motion of an impurity through the condensate at zero
temperature by considering the perturbation of a stationary solution of the
GP equation. We calculated the induced mass which contributes to the mass of
normal component. We find that the motion at small velocities is
dissipationless in one-, two-, and three- dimensional systems, although
movement with velocities larger than the speed of sound leads to a non-zero
drag force due to Cherenkov radiation of phonons. The expressions for the
drag force are calculated. We used results for the dynamic form factor of
exact Lieb-Liniger theory to investigate the velocity dependence of the drag
force in a 1D system. The form factor was calculated with the help of the
Haldane method of calculations of correlation functions. The drag force
exists at an arbitrarily small velocity of motion, but is very small in the
mean-field limit. The dynamic form-factor can be also directly measured by
applying a harmonic time-dependent perturbation on one-dimensional
condensates \cite{1DA,1DB}.

Acknowledgments: We thank T. Esslinger, F.~D.~M.~Haldane, B.~Jackson,
Yu.~Kagan, V.~Korepin, E.~Sonin and S.~Stringari for useful discussions.


\begin{references}
\bibitem{hom}  Considering an homogenious condensate, we assume that its
size is large enough, particularly that the gas is the Thomas-Fermi
conditions. Then results for a homogenious gas will be approximately valid
for a gas in a trap.
\end{references}
\end{document}